\documentclass[graybox]{svmult}

\usepackage{braket}

\usepackage{type1cm}  
\usepackage{makeidx}         
\usepackage{graphicx}        
\usepackage{multicol}        
\usepackage[bottom]{footmisc}

\newcommand{\ZT}[1]{\textquotedblleft#1\textquotedblright}%

\newcommand{\Tr}{\operatorname{Tr}}%

\newcommand{\ii}{\mathrm{i}}%

\usepackage[numbers]{natbib}
\usepackage{newtxtext}       %
\usepackage[varvw]{newtxmath}       

\makeindex             

\begin{document}

\title*{An introduction to reservoir computing}
\author{Michael te Vrugt}
\institute{Michael te Vrugt \at Institut f\"ur Physik, Johannes Gutenberg-Universit\"at Mainz, 55128 Mainz, Germany, \email{tevrugtm@uni-mainz.de}}
%
%
\maketitle

\textit{This book chapter will appear in: M. te Vrugt (Ed.), Artificial Intelligence and Intelligent Matter, Springer, Cham (2025)}

\hspace*{1cm}

\abstract{There is a growing interest in the development of artificial neural networks that are implemented in a physical system. A major challenge in this context is that these networks are difficult to train since training here would require a change of physical parameters rather than simply of coefficients in a computer program. For this reason, \textit{reservoir computing}, where one employs high-dimensional recurrent networks and trains only the final layer, is widely used in this context. In this chapter, I introduce the basic concepts of reservoir computing. Moreover, I present some important physical implementations coming from electronics, photonics, spintronics, mechanics, and biology. Finally, I provide a brief discussion of quantum reservoir computing.}

\section{Introduction}
You know that it is possible to perform machine learning tasks on a computer, but did you know that it is also possible to do so with a bucket of water? Precisely that was demonstrated in Ref.\ \cite{FernandoS2003}. Input data was mechanically fed into a bucket, recordings of the water surface then could be used for classification tasks. This is a form of \textit{reservoir computing} (RC), which this chapter will provide an introduction to.

\begin{figure}
	\includegraphics[scale=0.5]{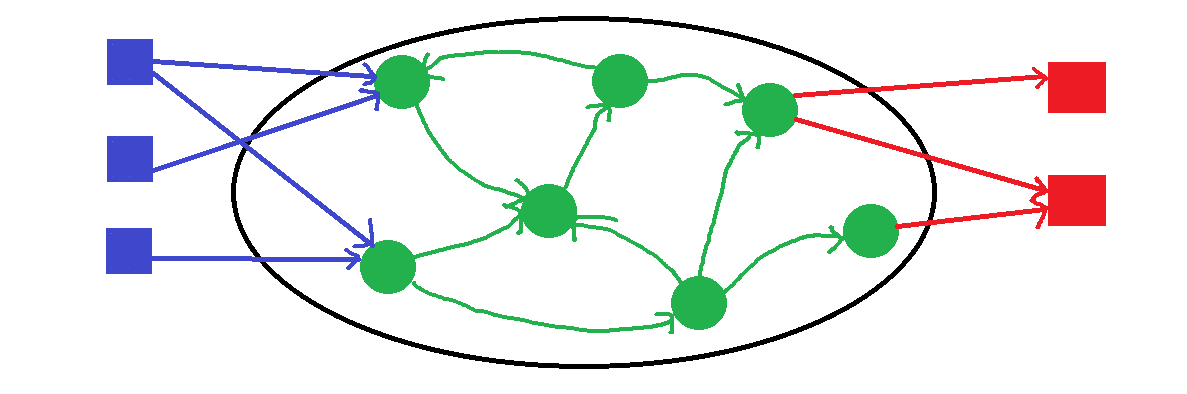}
	\caption{Illustration of the basic framework of reservoir computing. The input (blue) is fed into a recurrent neural network (green) and subsequently into an output layer (red). (A similar figure can be found in Ref.\ \cite{vanderSandeBS2017}.)}
	\label{fig1}
\end{figure}

The core idea of RC is that a significant portion of the computing task is performed not by a trained network, but by a very high-dimensional system (reservoir) that is essentially treated as a black box and whose output is fed into a single readout layer, which is the only component of the network that is trained. This setup is illustrated in Fig.\ \ref{fig1}. Since physical systems are generally much more difficult to train than neural networks on a computer, RC is a very promising approach for implementing artificial intelligence in a physical system (for example a bucket of water or -- more relevant in practice -- optical or magnetic systems). Consequently, RC has attracted considerable interest among physicists in recent years. From a computer science point of view, on the other hand, RC is useful for the otherwise difficult task of training recurrent neural networks.

The foundations of what is nowadays referred to as reservoir computing were independently developed by \citet{Jaeger2001}, who referred to it as \textit{echo state network}, and by \citet{MaassNM2002}, who called it \textit{liquid state machine}. (There were some earlier ideas in this direction -- see Refs.\ \cite{KirbyD1990,Kirby1991,Schomaker1991} -- that are largely unknown today \cite{NakajimaF2021}.) The name \ZT{reservoir computing} was coined in Refs.\ \cite{VerstraetenSS2005,VerstraetenSdS2007} to unify these concepts. RCs are particularly useful for tasks involving some temporal dynamics without needing external memory, since what they do depends also on the value of the input signal at previous times. Reviews of RC can be found in Refs.\ \cite{LukovsevivciusJ2009,vanderSandeBS2017,CucchiACBK2022,LukovseviciusJS2012,EverschorMWM2024}, a book on the topic was edited by \citet{NakajimaF2021}.

\section{Basic concepts of reservoir computing}

\subsection{\label{hiw}How reservoir computing works}
The discussion follows Refs.\ \cite{CucchiACBK2022,MujalMNGSZ2021} (and also adapts the notation used there).

Suppose that we have an $n$-dimensional training input signal $\vec{u}_\mathrm{train}(t)$ (depending on time $t$) that is supposed to lead to a certain $m$-dimensional output signal $\vec{y}_\mathrm{train}(t)$. For instance, if the system is supposed to predict time series data, $\vec{u}_\mathrm{train}(t)$ would be the first part of a time series and $\vec{y}_\mathrm{train}(t)$ the second part (that we want to predict from the first part) \cite{CucchiACBK2022}. The input signal is used to drive a high-dimensional nonlinear dynamical system that is referred to as the \textit{reservoir}. 

Specifically, the state $\vec{X}$ of the reservoir at time $t_i$ takes the form \cite{MujalMNGSZ2021}
\begin{equation}
\vec{X}(t_i)=\vec{f}(\vec{X}(t_{i-1}),\vec{u}(t_i))
\label{dynsyst}
\end{equation}
with the input signal $\vec{u}$ and a function $\vec{f}$. A sufficiently large subset of the state variables, summarized in a vector $\vec{x}(t)$, must be accessible. Finally, a readout function $\vec{F}$ maps the vector $\vec{x}(t)$ to the output $\vec{y}(t)$, i.e., 
\begin{equation}
\vec{F}(\vec{x}(t)) = \vec{y}(t).
\end{equation}
This readout function is the only thing that is touched during the training process. It is designed to minimize a loss function that typically that depends on the difference between the actual output $\vec{y}(t)$ and the desired output $\vec{y}_\mathrm{train}(t)$. In most cases, $\vec{F}$ is obtained via linear regression. Once $\vec{F}$ is known, one can apply the system to new input signals $\vec{u}(t)$.

This strategy has a number of advantages:
\begin{itemize}
    \item The training usually only consists in linear regression, which is simple, computationally inexpensive, and easy to implement.
    \item One only needs to train the readout function $\vec{F}$ and not the entire system. This is helpful if we are dealing with a physical system where the precise interactions between the parts are difficult to modify and perhaps not even fully known.
    \item One can use the same reservoir for different computing tasks by simply using different readout functions $\vec{F}$.
    \item It is a good approach for dealing with time series data.
\end{itemize}

Almost every physical system can be described by an equation of the form \eqref{dynsyst}, and therefore a large variety of physical system can be used as reservoirs (although in practice there are restrictions, see Section \ref{goodboi}). A notable feature of RCs is their temporal dynamics. Iterating Eq.\ \eqref{dynsyst} gives \cite{MujalMNGSZ2021}
\begin{equation}
\vec{X}(t_i)=\vec{f}(\vec{f}(\vec{f}(\vec{X}(t_{i-3}),\vec{u}(t_{i-2})),\vec{u}(t_{i-1})),\vec{u}(t_i)),
\end{equation}
showing that the system's state at a certain time depends on the system states and inputs at previous times. This allows the system to react to temporal input signals, such as spoken language. 

\subsection{Recurrent neural networks}
An essential distinction in this context (see also the chapter by Tobias Wand in this volume) is that between a \textit{feed-forward neural network} (FNN) and a \textit{recurrent neural network} (RNN). In a FNN, signals propagate in one direction - the first layer activates the second layer, the second layer the third layer and so on. In a RNN, on the other hand, information can also \ZT{move backwards}. This allows the system to have memory: If a system is supposed to process a signal $\vec{u}(t)$, then reacting to $\vec{u}(t_2)$ might require also information about $\vec{u}(t_1)$ with $t_1 < t_2$. This can be achieved by feeding back information about previous states of nodes. This distinguishes RC from the related concept of an \textit{extreme learning machine} \cite{HangHSY2015,WangLWZ2022}, which employs FNNs and do therefore not have memory of this type \cite{ButcherVSDH2013}.

The existence of closed loops implies in particular that RNNs can have a temporal dynamics even in the absence of external inputs. Therefore, while FNNs are (mathematically speaking) \textit{functions} -- they map a certain input to a certain output -- RNNs are (mathematically speaking) \textit{dynamical systems} \cite{LukovsevivciusJ2009}. This already indicates how they might be related to RC, which, as discussed in Section \ref{hiw}, is based on dynamical systems. Frequently, in computer science applications, the dynamical system constituting the reservoir is a RNN. 

In general, RNNs are difficult to train. FNNs are typically trained via gradient descent methods, where the parameters giving the connection weights are gradually changed to move the network's output closer to the target output. Gradient descent methods are also frequently, and in many different forms, applied to RNNs \cite{AtiyaP2000}. However, such methods are considerably more difficult to apply in the context of RNNs. Reasons for this include that a gradual parameter change might lead to a bifurcation that spoils convergence and that training times are very long since a single parameter change requires running the temperal dynamics for some time. RC was proposed as a new method for training RNNs that allows to avoid these problems by changing the weights only in a non-recurrent readout layer. The RNN itself can be created randomly and is unchanged during the training \cite{LukovsevivciusJ2009}. 

\subsection{Why does this work?}
At first sight, the idea of RC is somewhat counterintuitive. While it is of course easier to train only the final layer of a neural network rather than the entire one, there is certainly a reason why one usually trains all layers. Usually one would not get away with only changing the final layer, so why is it possible here?

\begin{figure}
\includegraphics[scale=0.5]{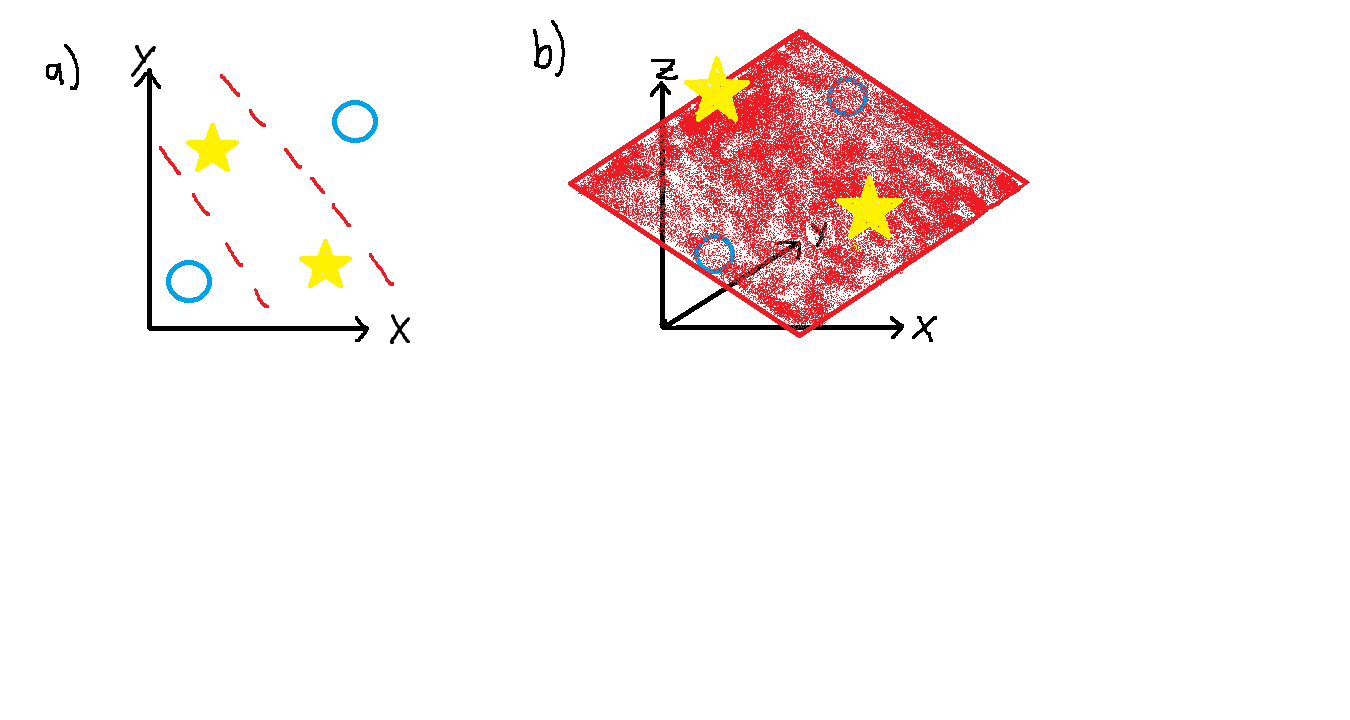}
\caption{Projection onto a high-dimensional space allows for separability. (A similar figure can be found in Ref.\ \cite{AppeltantEtAl2011}.)}
\label{fig2}
\end{figure}

What is exploited here is the high dimensionality of the reservoir. The information that we are interested in is in principle encoded in the input signal, but it is mixed up, nonlinearly, with a huge amount of other stuff that we are not interested in. The projection onto a higher-dimensional space allows for separability. This idea is illustrated in Fig.\ \ref{fig2}. Suppose we want to classify some inputs into stars and circles. Unfortunately, due to the way the data is distributed, it does not exhibit linear separability, i.e., we cannot find a hyperplane (in this case a line) that gives us the border between a region with stars and a region with circles (cf. Fig.\ \ref{fig2}a). If, however, we project the data onto a three-dimensional space (cf. Fig.\ \ref{fig2}b), we \textit{can} find a hyperplane (in this case a plane) that separates stars and circles. The separation problem has thereby become much simpler, and can be solved already by a single layer that is trained via linear regression. In RC, feeding the data into the reservoir corresponds to the projection from the two- to the three-dimensional plane (usually more dimensions will be involved) and the training of the readout layer corresponds to finding the plane \cite{Seoane2019}.

\section{\label{goodboi}What is a good reservoir?}
While the procedure discussed above can in principle be applied very generally, not every dynamical system does in practice make a good reservoir. A number of criteria have been developed that a reservoir should satisfy to be useful in this context -- although the broad range of systems that have been used for this purpose (consider, for example, the water bucket mentioned above) suggests that in practice these criteria are quite flexible \cite{Seoane2019}.

First, \textit{reproducibility} is of course important. If two input signals are very similar, the output signals should be similar as well. This is related to the ability of the system to generalize from the data it has been trained on to other input data \cite{Seoane2019}. Second, if the input signals are sufficiently different, the output signals should also be different (\textit{separation}). This determines the reservoir's ability to distinguish different sorts of inputs. A reservoir computer that always produces the same output regardless of the input obviously satisfies the reproducibility property, but would not be really useful.

Moreover, \textit{fading memory} is an important property of reservoir computers \cite{CucchiACBK2022}. The computer is supposed to process a time series $\vec{u}(t)$, not just he instantaneous value of $\vec{u}$ (the latter would not require a RNN). Therefore, it needs to have memory. On the other hand, the current value of $\vec{u}$ should still be what is most important, values at earlier times become gradually less important. In other words, the memory should gradually fade. The computer should relax to a quiescent state if there is no external input \cite{vanderSandeBS2017}, and its behavior should not depend on the initial condition $\vec{X}(0)$ of the reservoir. This is referred to as the \textit{echo state property} - the influence of initial conditions should gradually vanish \cite{YildizJK2012}. The timescale of the fading memory should be comparable to the timescales of the input signal \cite{KasparRvdWWP2021}. Since it depends on the application what the timescale of the input signal is, the adequate reservoir system may differ depending on the application.

Also, it is often considered advantageous to operate the reservoir computer in a parameter range close to an instability (for example, in the vicinity of a transition to chaos), since there its behavior is particularly complex and therefore has a very high computational complexity. Intuitively, the reason is that such regimes present a compromise between ordered phases (where the behavior is stable and thus reliable, but where initial conditions are also quickly erased by the approach to an attractor making the system less able to react to an input) and chaotic phases (where the system's behavior is strongly affected by small differences in the initial conditions, harming separability) \cite{Seoane2019}. There are, however, also counterexamples, i.e., systems where moving close to an instability actually decreases the performance of a reservoir computer \cite{Carroll2020}. In fact, Herbert Jaeger (in his foreword for Ref.\ \cite{NakajimaF2021}) argues that the idea that RC should operate close to chaos or criticality is a \ZT{myth}, neither mathematically well-defined nor empirically confirmed.

\section{Physical reservoir computing}
The discussions so far were concerned with general concepts in artificial intelligence. This book, however, has a specific focus on the relation of AI to physics, and there is a reason that RC features so prominently in the introductory part of this book. This reason is \textit{physical reservoir computing}.

In physical RC, the reservoir is a physical system. Thereby, physical RC allows to perform machine learning tasks using physical systems. There has, in recent years, been a considerable amount of work on the development of experimental setups that can be used for this purpose. In this section, I will briefly discuss some examples. Later chapters will address specific such as in optics (in the chapter by Kathy L\"udge/Lina Jaurigue), spintronics (in the chapter by Atreya Majumdar/Karin Everschor-Sitte), and soft matter (in the chapter by Julian Jeggle/Raphael Wittkowski). The classification in different types of reservoirs (electronic, photonic, spintronic, mechanical, biological, chemical) is adapted from Ref.\ \cite{TanakaYHNKTNNH2019}. Note that this list is not exhaustive.

\subsection{Electronic reservoir computing}
Standard computers are based on electronics, and therefore it is a very natural idea to use electronic systems. This has been achieved in a broad number of ways (see Refs.\ \cite{TanakaYHNKTNNH2019,LiangTZGW2024} for an overview), of which I discuss here just one, namely \textit{memristors}. A memristor is a resistor that possesses memory, i.e., whose resistance changes based on the current that has passed through it. Memristors are therefore useful devices in applications like RC where memory is important. They can be used to mimic the plasticity of biological neurons, and allow for nonlinear transformations of input signals \cite{TanakaYHNKTNNH2019}. A memristor-based RC was first proposed by \citet{KulkarniT2012}.

\subsection{Photonic reservoir computing}
In \textit{photonics}, information processing is based not (solely) on electric currents, but on light (photons). RC has developed into a widely used approach in photonics, see Ref.\ \cite{vanderSandeBS2017} for a review (I am loosely following this reference here). An important approach is the implementation of RC in on-chip-photonics, where photonic systems are integrated into chips. This allows the systems to be produced and sold on industrial scales and to therefore use them for high-speed low-power-consumption computing. This approach was theoretically suggested in Ref.\ \cite{VandoorneDSVBBV2008} and later realized in hardware \cite{VandoorneEtAl2014}. 

A further interesting approach is \textit{delay reservoir computing} \cite{HulserKJL2022}. Delay systems have been of considerable interest for optics in the past years \cite{SeidelGJ2022,KochSJG2023}. They are described mathematically by delay differential equations, which differ considerably from ordinary differential equations since they possess an infinite-dimensional phase space -- for solving it, one needs to specify not only the state of the system at a single initial time, but on an entire time interval. Using delay systems, one can therefore achieve a high-dimensional phase space (which is advantageous in the context of RC) even with a very simple setup \cite{vanderSandeBS2017}.

Photonic approaches to neuromorphic computing are discussed further in the chapters by Kathy L\"udge/Lina Jaurigue and by Lennart Meyer/Rongyang Xu/Wolfram Pernice in this book.

\subsection{Spintronic reservoir computing}
\textit{Spintronics} is a field of technology where information processing is based not only on the electric charges (as in electronics), but also on the spins (elementary magnetic moments) of electrons. Spintronics has become increasingly popular in neuromorphic computing in general and RC in particular, see Refs.\ \cite{ZhouC2021,FinocchioDCEAZ2021,GrollierQCEFS2020,EverschorMWM2024,YanHBTLS2024,LiangTZGW2024} for reviews. Spintronic systems can be used to build artificial synapses, thereby mimicking the structure and functionality of biological brains \cite{GrollierQCEFS2020}. An introduction is provided in the chapter by Atreya Majumdar/Karin Everschor-Sitte in this book.

An interesting recent proposal in this context is \textit{Brownian reservoir computing} based on skyrmions \cite{RaabBBDRKMK2022,BremsKV2021,BremsRVK2023,BremsEtAl2023}. Brownian motion \cite{Brown1828} is the random thermal motion of particles, which is a central phenomenon in soft matter physics, but also arises in magnetic systems. An example are magnetic skyrmions, which are whirl-like topological magnetic nanostructures that have particle-like diffusion behavior reminiscent of neurotransmitters and can be used as information carriers in spintronics \cite{GrollierQCEFS2020}. In Brownian computing, one employs thermal fluctuations -- which in most systems are present anyway -- for computing purposes to achieve a high energy efficiency. It is of course helpful if the employed Brownian system can be easily integrated into a computer, which is why magnetic nanosystems such as skyrmions are useful here \cite{BremsRVK2023}. Brownian RC based on skyrmions was realized experimentally by \citet{RaabBBDRKMK2022}, who demonstrated that this approach is very promising for energy-efficient computing.

\subsection{Mechanical reservoir computing}
Mechanical systems can make for useful reservoirs. Robotic systems, in particular from soft robotics (where the bodies of the robots are flexible), have been repeatedly used in this context. \citet{HauserIFP2011} have modeled this using the example of a nonlinear mass-spring-damper system connected to a mechanical network, which was intended to represent in a simple way the body of a soft robot (or biological system) and which exhibits the complex nonlinear dynamics required for successful RC. \citet{NakajimaLHP2014} employed a silicon-based robot arm inspired by the arm of an octopus, with the input being the rotation of the arm and the output being measured strain. While the noisy and nonlinear dynamics of soft robots is often perceived as disadvantageous, it can be very useful in the context of RC. (This paragraph follows Ref.\ \cite{Hauser2021}.)

\subsection{Biological and chemical reservoir computing}
Reservoir computing has always had a close connection to neurobiology. In particular, work on RC has been motivated by attempts to understand information processing in mammalian brains \cite{SumiEtAl2023}. For instance, it has been proposed that the cerebellum might work like a liquid state machine \cite{YamazakiT2007}, and experiments on mice \cite{CazettesMMMARM2023} suggest that the mouse brain exploits principles of RC \cite{CazettesMMMARM2023,LiangTZGW2024}. Moreover, RC -- which is frequently based on random neural networks and noisy systems -- might explain why the brain works so accurately despite being a rather noisy system \cite{LukovsevivciusJ2009}. It is therefore a promising direction of work to use biological neural networks for RC tasks \cite{SumiEtAl2023}. Biological neurons are in a sense the most obvious, but not the only approach to biological RC. For example, it has been proposed to realize RC based on Escherichia coli bacteria \cite{JonesRF2007}. Another variant, namely DNA reservoir computing \cite{GoudarziLS2013}, will be discussed in the chapter on DNA neural networks in this book. This approach is based on employing chemical systems for RC, an idea that is also used in non-biological contexts (for example based on electrolyte solutions \cite{KanNA2021}). See the chapter by Julian Jeggle/Raphael Wittkowski in this volume for a discussion of the related concept of active matter RC.

\section{Quantum reservoir computing}
In the wake of the currently growing interest in quantum computing in general and quantum-mechanical approaches to machine learning in particular, \textit{quantum reservoir computing} \cite{FujiiN2017,MartinezGNSZ2021,FujiiN2021,SuzukiGPYY2022,MujalMNGSZ2021} has attracted some interest. Here, one employs quantum-mechanical reservoirs and thereby aims to exploit the advantages of quantum computers for RC. In this section, I will introduce the elementary ideas of how this works. The discussion follows Ref.\ \cite{FujiiN2017}, which was one of the first articles on this topic. A more general introduction to quantum machine learning can be found in the chapter by Ivana Nikoloska in this volume.

Quantum states are represented by vectors in complex Hilbert states. In the context of quantum computing, the minimal information unit is a \textit{qubit}, corresponding to a two-dimensional complex vector in a vector space spanned by the vectors $\ket{0}$ and $\ket{1}$. In general, the state of a system of $N$ qubits is described by a $2^N x 2^N$ Hermitian matrix $\rho$, the \textit{density matrix}. The quantum system is said to be in a \textit{pure state} if $\rho$ can be written as $\rho = \ket{\psi}\bra{\psi}$, where $\ket{\psi}$ is a $2^N$-dimensional vector and $\bra{\psi}$ is a covector to $\ket{\psi}$. (For instance, if $\ket{\psi} = (1,2,3)^\mathrm{T}$, then $\bra{\psi} = (1,2,3)$.) If the density matrix at time $t$ is $\rho(t)$, then the density matrix at time $t+\tau$ is
\begin{equation}
\rho(t+\tau) = e^{-\ii H \tau}\rho(t)e^{\ii H \tau}
\label{timeevolution}
\end{equation}
with the Hamiltonian $H$ (a $2^N x 2^N$ Hermitian matrix that determines the dynamics and whose eigenvalues correspond to the energy levels of the quantum system). For an arbitrary observable $A_i$, which is also represented by a $2^N x 2^N$ Hermitian matrix, the expectation value is given by
\begin{equation}
a_i(t) = \Tr(\rho(t)A_i)
\label{expectationvalue}
\end{equation}
with the trace $\Tr$.

What is now required is a way to feed an input signal $\vec{u}(t)$ into the system and to get an output signal $\vec{x}(t)$ that can then be fed into the readout function $\vec{F}$. Let us consider for simplicity a one-dimensional input signal $u(t)$, which we sample in $M$ discrete time intervals of length $\tau$ to get a sequence $\{u_k\}$ with $u_k = u(k\tau)$ and $k = 0, 1, ... M$. At each time $k\tau$, the state of the first qubit is changed to $\rho_{u_k} = \ket{\psi_{u_k}}\bra{\psi_{u_k}}$ with
\begin{equation}
\ket{\psi_{u_k}}= \sqrt{1 - u_k}\ket{0} + \sqrt{u_k}\ket{1}.
\end{equation}
The density matrix $\rho$ is thereby replaced by
\begin{equation}
\rho_{u_k}\otimes \Tr_1(\rho),
\end{equation}
where $\otimes$ is a tensor product and $\Tr_1$ is a trace over the degrees of freedom of the first qubit. Afterwards, the density matrix is time evolved via Eq.\ \eqref{timeevolution} for a time $\tau$. This time has to be optimized in order to optimize the performance of the computer (see Ref.\ \cite{FujiiN2017}). For the output $\vec{x}(t)$, we then pick some observables $A_i$ and assemble them in a vector $\vec{A}$. Then, we can obtain the output vector from their expectation values as
\begin{equation}
\vec{x}(t) = \Tr(\rho(t)\vec{A}).
\end{equation}
Specifically, \citet{FujiiN2017} choose $A_i$ as the Pauli operator acting on the $i$th qubit.

The dimension of the quantum-mechanical Hilbert space increases exponentially with the number of qubits $N$, giving rise to an exponentially increasing number of nodes in the reservoir. For readout purposes, these are split into \textit{true nodes} (the observed ones) and \textit{hidden nodes} (the rest). The signals are sampled not only at the time $k\tau$, but also at several times in between. Dividing the time interval into $V$ parts gives rise to $V$ \textit{virtual nodes} and allows to increase the number of nodes from $N$ to $NV$ via temporal multiplexing. Thereby, the exponentially large Hilbert space is monitored via a polynomial number of signals. This is the distinguishing feature of quantum RC compared to other RC approaches \cite{FujiiN2017}. Changing $\tau$ corresponds to a change of the dynamics of the reservoir, whereas changing $V$ corresponds to a change of the way it is observed \cite{NakajimaFNMK2019}.

An important feature of quantum systems is also the way in which they interact with the environment. Such interactions lead to dissipation and decoherence \cite{SanniaMSGZ2022}, where quantum states are destroyed by interactions with the environment. Moreover, performing a measurement of a quantum state generally changes it, a phenomenon giving rise to the famous quantum measurement problem \cite{FriebeKLNPS2018}. Usually, interactions with the environment are not beneficial for the performance of quantum computers. One can, however, also try and exploit such effects in quantum reservoir computing, as has recently been demonstrated for both measurements \cite{MujalGSZ2023} and dissipation \cite{SanniaMSGZ2022}. 

\section{Outlook: Relation to intelligent matter}
If we loosely understand \ZT{intelligent matter} as \ZT{physical materials perform tasks similar to those expected from computer systems that we would refer to as (artificially) intelligent}, then RC seems to be, if not an instance of it, then at least an important step towards it. We have here physical systems that can be employed in computational tasks of the form that appear in machine learning.

Nevertheless, according to \citet{KasparRvdWWP2021}, reservoir computing systems in the form described here do not constitute \ZT{intelligent matter} in the technical sense:
\begin{itemize}
    \item The systems possess fading memory, whereas intelligent matter needs to have long-term memory.
    \item The readout function $\vec{F}$ still needs to be trained manually, the system does not adapt on its own.
\end{itemize}
Regarding the first point, it should be noted, however, that the fading memory can be tuned to fade rather slowly if this is desired in a certain context.

RC does nevertheless have significant potential for the development of \ZT{true} intelligent matter, in particular when considering its relation to evolutionary dynamics (a topic reviewed in Ref.\ \cite{Seoane2019}). After all, RC is a possible working principle of biological brains. It is conceivable that RC emerges in evolutionary contexts, as it has certain advantages (such as the low cost of learning and the fact that external sytems can be used to carry out computations) that could give biological systems exploiting this paradigm a fitness advantage. An evolutionary evolving RC system would be a system that evolves its computing capabilities in adaptation to the environment, bringing it closer to actual intelligence. A possible disadvantage of RC in evolutionary contexts, \citet{Seoane2019} suggests, is that (since the reservoir needs to be high-dimensional), it requires systems to perform a lot of activity that is not really used for computing, making it energetically costly (which leads to a fitness disadvantage). A fine-tuned neural network can have a smaller number of nodes.

\section{Summary}
In this chapter, I have introduced the basic ideas of reservoir computing. Here, one uses a very high-dimensional recurrent neural network and trains only the final layer. This makes it possible to use for the rest of the network a physical system whose properties might be difficult to tune or not fully known. A variety of systems have been used here, ranging from buckets of water to optical and magnetic setups. Reservoir computing is a very promising tool for implementing artificial intelligence in nanosystems, and will continue to be a thriving field of research in the coming years. 

\section{Acknowledgements}
I thank Raphael Wittkowski for very helpful discussions on this topic. This work was funded by the Deutsche Forschungsgemeinschaft (DFG, German Research Foundation) in the framework of SFB 1551; Project No. 464588647 and SFB 1552; Project No. 465145163.

\end{document}